\documentclass[sigconf,nonacm]{acmart}
\usepackage{booktabs}
\usepackage{microtype}
\usepackage{xspace}
\usepackage{xcolor}
\newcommand{\SearchUniverseN}{300}

\newcommand{\CorpusN}{173}

\newcommand{\TopHundredEligible}{100}
\newcommand{\TopHundredExposed}{4}
\newcommand{\TopHundredDurable}{54}
\newcommand{\TopHundredOverallPct}{4.0}
\newcommand{\TopHundredPct}{7.4}

\newcommand{\ResolveP}{4.6}
\newcommand{\MergeP}{88.2}
\newcommand{\ManifestBytes}{973}
\newcommand{\CheckpointDelta}{2978}
\newcommand{\BoundResources}{6}
\newcommand{\BranchRecords}{2}

\newcommand{\system}{\textsc{SemIso}\xspace}
\newcommand{\ph}[1]{\par\kern3.75pt\noindent\textbf{#1}\kern0.18em\textemdash\kern-0.42em\textemdash\kern0.22em}
\renewcommand\footnotetextcopyrightpermission[1]{}
\AtBeginDocument{\pagestyle{plain}}
\title{BEGIN AI TRANSACTION: \\ Semantic Isolation for Durable AI Workflows}
\author{Barzan Mozafari}
\affiliation{\institution{University of Michigan}\city{Ann Arbor}\state{Michigan}\country{USA}}
\email{mozafari@umich.edu}
\begin{document}
\begin{abstract}

An AI execution can now outlive the environment in which it began. What once fit inside one model call increasingly unfolds across pauses, retries, branches, subagents, and model-selected tools. Meanwhile, prompts, model aliases, indexes, policies, and tools are deployed independently: stable names can acquire new behavior, and workflows can discover resources only after they start.
The workflow can therefore combine saved state with changed assumptions, producing an internally inconsistent result even when every call succeeds.
This is an isolation problem: database transactions constrain concurrent data updates, but workflow checkpointing provides no corresponding contract for concurrent changes to an AI workflow's semantic environment.

We define four automatically detectable anomalies---semantic read skew, compatibility skew, context escape, and merge skew.
To control which anomalies are allowed, we derive a partial order of isolation levels, from Semantic Read Committed to Semantic Snapshot Isolation, by combining three independent guarantees: resource stability, cross-resource compatibility, and continuation inheritance.
In a conservative source audit of the 100 most-starred public repositories with executable LangGraph code,
we find that \TopHundredPct\%
of codebases with durable workflows resolve live or dynamically selected semantic resources within the same workflow, without an evident immutable binding.
We show that these guarantees can be checked and enforced efficiently in middleware. Our prototype, \system,
propagates semantic context and blocks incompatible resources and branch merges with microsecond-scale checks.
\end{abstract}
\maketitle

\section{Introduction}
\label{sec:intro}

The unit of AI computation is expanding from a single model invocation into a durable execution that can pause for a person, retry after failure, split into parallel branches, delegate to subagents, and discover tools as it runs. At the same time, consequential application logic is moving outside the program binary: prompts, model aliases, retrieval indexes, policies, and tool descriptions are deployed independently and can change behind stable names. These resources also depend on one another: for example, a vector index is meaningful only with a compatible embedding model, even though an incompatible query may remain type-correct and execute successfully.
Thus, a long-running workflow can retain its data and control position while losing the assumptions that give that state meaning.
Just as database isolation controls which concurrent updates a transaction may observe, semantic isolation should control which concurrent changes to semantic resources (e.g., prompts, models, indexes, policies, and tools) an AI workflow may observe.

\begin{figure*}[t]
\centering
\begin{minipage}[t]{.485\textwidth}
\textbf{(a) Semantic read skew}\par\smallskip
\includegraphics[width=\linewidth]{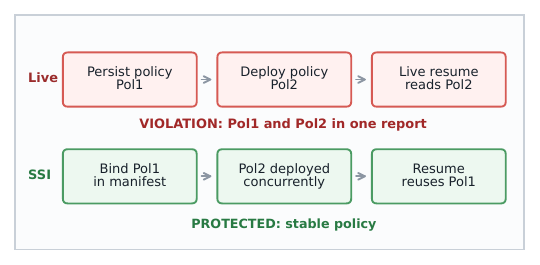}
\end{minipage}\hfill
\begin{minipage}[t]{.485\textwidth}
\textbf{(b) Compatibility skew}\par\smallskip
\includegraphics[width=\linewidth]{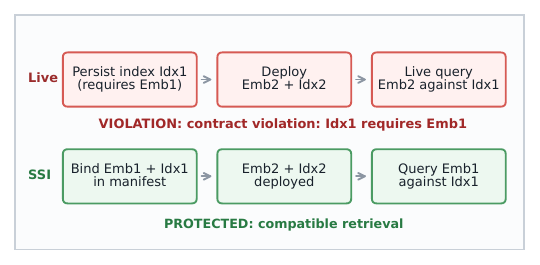}
\end{minipage}

\smallskip
\begin{minipage}[t]{.485\textwidth}
\textbf{(c) Context escape}\par\smallskip
\includegraphics[width=\linewidth]{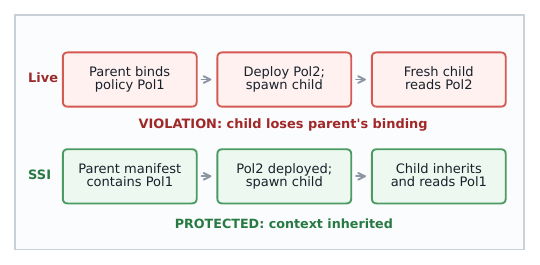}
\end{minipage}\hfill
\begin{minipage}[t]{.485\textwidth}
\textbf{(d) Merge skew}\par\smallskip
\includegraphics[width=\linewidth]{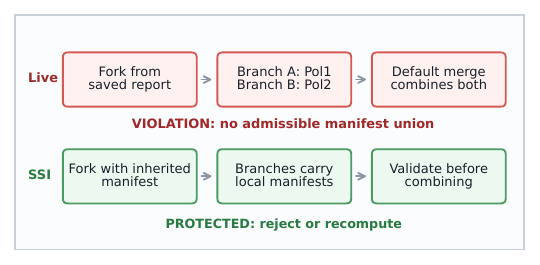}
\end{minipage}
\Description{Four panels show a prompt changing after a pause, an embedding model used with the wrong index, a child losing its parent's context, and incompatible branch outputs being merged.}
\caption{An example illustrating the four semantic-isolation anomalies: a compliance workflow retrieves vendor evidence, pauses for approval, and resumes after deployments. Red shows live resolution; green shows Semantic Snapshot Isolation (SSI). The notation matches Section~\ref{sec:semantic-state}: $\mathsf{Pol}_1/\mathsf{Pol}_2$ are policy versions, $\mathsf{Emb}_1/\mathsf{Emb}_2$ are embedding models, and $\mathsf{Idx}_1/\mathsf{Idx}_2$ are vector indexes.
Each panel isolates one failure and its protected outcome.}
\label{fig:history}
\end{figure*}

\ph{One report, incompatible assumptions} Consider a compliance agent that retrieves vendor evidence under one policy, saves its findings, and pauses for human approval. While it waits, operators publish a stricter policy, rebuild the search index for a new embedding model, and change what a risk score means behind the same tool name. After approval, one branch resumes with the saved findings while another uses the new resources. Every call succeeds, but the agent can retrieve the wrong evidence, interpret a score backward, and combine recommendations written under incompatible policies. The final report has no coherent semantic basis even if every database read used snapshot isolation. Figure~\ref{fig:history} separates the four ways this execution can fail.

\ph{Deployments as concurrent semantic writes}
Transactions spare database applications from reasoning about every possible interleaving: they define a unit of execution, while isolation levels specify which effects of concurrent updates that unit may observe~\cite{berenson95}. AI workflows need the same kind of choice for \emph{semantic resources}---prompts, models, indexes, policies, and tools that determine what saved state means and how the next step interprets it. We model each deployment as a concurrent semantic write relative to every active workflow.

This gap can be illustrated in LangGraph, a widely adopted open-source framework for building stateful, durable AI-agent workflows~\cite{langgraphrepo}.
 LangGraph checkpoints thread state for interruption and recovery, yet resumes do not automatically bind independently managed prompts, indexes, or tools~\cite{langgraphrepo,langgraphpersist}. In our source audit, \TopHundredPct\% of durable workflows among the 100 most-starred executable LangGraph codebases combine persisted state with live or dynamically selected semantic resources. Such a workflow can therefore resume with saved evidence or decisions whose meaning has changed because a prompt, index, or tool now resolves differently
 (see Section~\ref{sec:evaluation}).

\ph{Why narrower mechanisms stop short}
Existing mechanisms protect narrower boundaries. Provenance and registries record which versions were used~\cite{mlmd,mlflow}, but do not prevent an inconsistent commit. As previously mentioned,
LangGraph checkpointers save thread state~\cite{langgraphpersist}; its recommended \texttt{flow\_version} can retain old node logic after resume, but does not bind an external prompt, index, or tool~\cite{langgraphcompat}.
\emph{Static pinning} records the exact version of every dependency identity \emph{known at startup}; it therefore provides complete coverage when the dependency universe is closed~\cite{dvc,zenml}.
 A durable agent, however, may discover a tool or child only after model output or human input; predeclaring every candidate changes that assumption and can impose unnecessary retention. Timestamp-based snapshots~\cite{berenson95} also fall
  short: simultaneously available resources can be incompatible, while versions created at different times can remain compatible.

\begin{figure}[t]
\centering\includegraphics[width=\columnwidth]{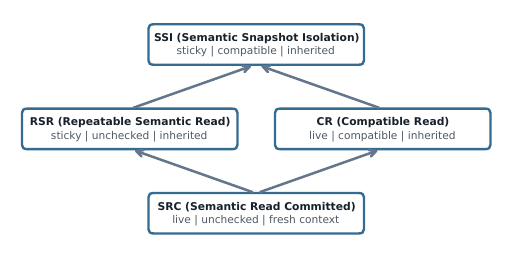}
\Description{A partial-order lattice places SRC below incomparable RSR and CR, both below SSI.}
\caption{The isolation levels form a partial order over three choices shown explicitly in each box:
(1) live versus sticky resource binding,
(2) unchecked versus compatibility-checked composition, and (3) fresh versus inherited continuation context. The two middle levels are incomparable: Repeatable Semantic Read adds sticky bindings, whereas Compatible Read adds compatibility checking. Semantic Snapshot Isolation provides both.}
\label{fig:isolation-lattice}
\end{figure}

\ph{From anomalies to isolation choices}
In this paper, we define semantic read, compatibility, context-escape, and merge skew, as summarized in Table~\ref{tab:anomaly-summary}.
For each anomaly, we specify a condition over resource versions, declared contracts, and inherited workflow context that a runtime can check automatically, without judging the generated answer's factual correctness or quality.
 The guarantees that prevent these anomalies are independent: stability S, compatibility C, and inheritance I therefore form a partial-order lattice (Figure~\ref{fig:isolation-lattice}) rather than a single strength ladder. From this lattice, we define four named isolation levels: Semantic Read Committed (SRC), Repeatable Semantic Read (RSR), Compatible Read (CR), and Semantic Snapshot Isolation (SSI) (Table~\ref{tab:levels}).

\begin{table}[t]
\centering\footnotesize
\setlength{\tabcolsep}{3pt}
\caption{Four independently checkable semantic anomalies.}
\label{tab:anomaly-summary}
\begin{tabular}{lp{.65\columnwidth}}
\toprule
Anomaly & Observable condition\\
\midrule
Read skew & One identity resolves to different versions.\\
Compatibility skew & Selected versions violate a declared contract.\\
Context escape & A continuation loses an inherited semantic obligation.\\
Merge skew & Locally valid branch manifests have no admissible union.\\
\bottomrule
\end{tabular}
\end{table}

\ph{Enforcing semantic isolation}
The isolation levels specify what a workflow may observe; enforcing those choices requires runtime support across pauses, resumes, and dynamically discovered resources. We therefore propose \system,   middleware   for stateful AI-workflow frameworks. The application selects an isolation level when invoking a workflow. \system then records resource bindings and compatibility requirements in a semantic manifest, checkpoints that manifest with workflow state, restores it on retry or resume, propagates it to branches and children, and validates late-discovered resources and branch combinations before use.

In this paper, we make the following contributions:
\begin{enumerate}
\item We formulate semantic isolation as a transactional problem for durable AI workflows and define four automatically decidable anomalies.
\item We derive a partial-order family of isolation levels
over resource stability, cross-resource compatibility, and continuation inheritance.
\item We provide real-world evidence for the problem: \TopHundredPct\% of durable workflows in public codebases analyzed in our experiment persist state with live or dynamic semantic resources, and real LangGraph/Qdrant executions independently reproduce all four anomalies.
\item We show that AI-workflow middleware can verify and enforce semantic isolation efficiently
by implementing \system with online manifest extension, continuation inheritance, and branch-merge validation (Section~\ref{sec:runtime}).
\end{enumerate}

The rest of the paper defines semantic state and its anomalies (Section~\ref{sec:semantic-state}), derives isolation levels (Section~\ref{sec:isolation-levels}), presents the runtime (Section~\ref{sec:runtime}), evaluates the motivating evidence and implementation (Section~\ref{sec:evaluation}), positions the work against related systems (Section~\ref{sec:related}), and concludes with   future work (Section~\ref{sec:conclusion}).

\section{Semantic State and Anomalies}

We now formalize the semantic state and the four failures introduced by the motivating example in Section~\ref{sec:intro} and illustrated in Figure~\ref{fig:history}. We first define the objects recorded during execution, then give the shortest history that demonstrates each anomaly and an automatically checkable condition for detecting it.

We model an immutable resource version as a tuple containing its resource identity, version identifier, type, content or contract hash, availability, dependencies, and compatibility attributes. We represent an execution trace as an ordered sequence of resolution, deployment, continuation, fork, and merge events. We write $r_T(x{:}v)$ when execution context $T$ resolves resource identity $x$ to immutable version $v$. We define $T$'s \emph{semantic manifest}, $M_T$, as the partial identity-to-version mapping accumulated by that context together with its inherited compatibility obligations.

We use the following notation in reference to the motivating example from Section~\ref{sec:intro}:
\begin{description}
\item[Policy:] Let $\mathsf{Pol}_1$ and $\mathsf{Pol}_2$ be two versions of one policy or prompt identity.
\item[Search:] Let $\mathsf{Emb}_1$ and $\mathsf{Emb}_2$ be two embedding versions. $\mathsf{Idx}_1$ is an index built for $\mathsf{Emb}_1$.
\item[Tools:] Let $\mathsf{Sch}_1$ be a model-visible tool schema and $\mathsf{Impl}_2$ be a tool implementation with a potentially different contract identifier.
\item[Contexts:] Let $T$ be a parent execution context, $c$ be a child, and $a,b$ be parallel branches.
\end{description}

\begin{sloppypar}\ph{Semantic read skew} The shortest history that demonstrates this anomaly is $r_T(\mathsf{Pol}{:}\mathsf{Pol}_1);$ \allowbreak$\mathit{deploy}(\mathsf{Pol}{:}\mathsf{Pol}_2);$ \allowbreak$\mathit{resume}(T);$ \allowbreak$r_T(\mathsf{Pol}{:}\mathsf{Pol}_2)$. One inherited context resolves the same identity to two immutable versions. We detect the anomaly when the trace contains $r_T(x{:}v_i)$ and $r_T(x{:}v_j)$ with $v_i\neq v_j$. Equivalent records can consequently be judged under different instructions.\end{sloppypar}

\begin{sloppypar}\ph{Compatibility skew} The shortest retrieval history that demonstrates this anomaly is $r_T(\mathsf{Emb}{:}\mathsf{Emb}_2);$ \allowbreak$r_T(\mathsf{Idx}{:}\mathsf{Idx}_1)$ although $\mathsf{Idx}_1$ declares a dependency on $\mathsf{Emb}_1$. A tool variant presents $\mathsf{Sch}_1$ to the model but invokes $\mathsf{Impl}_2$ with a different contract identifier. Every access can remain well typed and available. We detect the anomaly when at least one declared dependency or compatibility condition evaluates false over $M_T$.\end{sloppypar}

\begin{sloppypar}\ph{Context escape} The shortest history that demonstrates this anomaly is $r_T(\mathsf{Pol}{:}\mathsf{Pol}_1);$ \allowbreak$\mathit{spawn}_T(c);$ \allowbreak$\mathit{deploy}(\mathsf{Pol}{:}\mathsf{Pol}_2);$ \allowbreak$r_c(\mathsf{Pol}{:}\mathsf{Pol}_2)$, where $c$ should inherit $T$'s semantic context but starts fresh. Both versions are individually valid and compatible, so no compatibility condition fails. We detect the anomaly when a retry, resume, branch, child, or late resolver omits a profile or manifest obligation that its parent required it to inherit.\end{sloppypar}

\begin{sloppypar}\ph{Merge skew} The shortest history that demonstrates this anomaly is $\mathit{fork}_T(a,b);$ \allowbreak$r_a(\mathsf{Pol}{:}\mathsf{Pol}_1);$ \allowbreak$r_b(\mathsf{Pol}{:}\mathsf{Pol}_2);$ \allowbreak$\mathit{merge}_T(a,b)$. Each branch manifest is locally valid, but their union binds one identity to conflicting versions and the runtime nevertheless combines their artifacts. More generally, resources selected independently by the branches may violate a compatibility condition only after union. We detect the anomaly when $M_a$ and $M_b$ are each admissible, their union has no admissible common extension, and the branch artifacts are nevertheless merged.\end{sloppypar}

No anomaly implies another: each has a witness history in which the remaining three are absent.
Read skew needs only two versions of one identity in one inherited context. Compatibility skew can occur with one stable, correctly inherited manifest. Context escape can occur when parent and child each use a single, internally compatible version. Merge skew can occur when both branches are locally stable and compatible but their union is inadmissible.
We can therefore define and detect each anomaly independently.
Note that the same four predicates apply whether the relevant access occurs during ordinary execution, retry, resume, routing, child-agent execution, or tool invocation.

\ph{Isolation versus nondeterminism}
Isolation and model nondeterminism are orthogonal: an AI workflow may produce different outputs while satisfying all four anomaly predicates; conversely, it may violate a predicate even when two generated outputs happen to match.
Database isolation makes the same distinction: a history can be serializable even when a transaction performs nondeterministic internal computation, because serializability constrains the observable ordering of reads and writes rather than requiring repeated executions to produce identical results.
Similarly, semantic isolation constrains the resource versions and contracts an execution may observe, not the determinism of the computation that consumes them.

Having defined the failure modes, we next derive isolation guarantees that prevent selected subsets of them.

\label{sec:semantic-state}

\section{Isolation Levels for AI Transactions}
\label{sec:isolation-levels}

We encode an AI transaction's isolation contract as $\langle S,C,I\rangle$
where $S\in\{S0,S1\}$ is resource \emph{stability},
$C\in\{C0, C1\}$ is cross-resource \emph{compatibility},
and $I\in\{I0, I1\}$ is continuation \emph{inheritance}. Under $S0$, each access may resolve the currently published immutable version. Under $S1$, the first access binds identity $x$ to version $v$, and every later access in the inherited context must return $v$.
If $v$ is unavailable, preserving $S1$ requires aborting rather than silently substituting a version; an explicit capture/taint policy may instead weaken the guarantee. Under $C0$, published versions compose unchecked; under $C1$, their set must be dependency-closed and satisfy every declared compatibility predicate. Under $I0$, a continuation may start fresh; under $I1$, retries, resumes, forks, children, and late discovery inherit the profile and manifest obligations, and branches undergo merge admission.

\begin{table}[t]
\centering\footnotesize
\setlength{\tabcolsep}{2.2pt}
\caption{Named isolation levels. ``Allows'' lists anomaly classes not excluded by the level: semantic read skew (SR), compatibility skew (CS), context escape (CE), and merge skew (MS).}
\label{tab:levels}
\begin{tabular}{lllll}\toprule
Level & Profile & Allows & Main cost & Impl.\\\midrule
SRC & $\langle S0,C0,I0\rangle$ & SR/CS/CE/MS & fresh/live & yes \\
RSR & $\langle S1,C0,I1\rangle$ & CS & retention & yes \\
CR & $\langle S0,C1,I1\rangle$ & SR/MS & contracts & yes \\
SSI & $\langle S1,C1,I1\rangle$ & none & retention+admission & yes \\
\bottomrule\end{tabular}

\end{table}

\begin{sloppypar}\ph{A lattice, not a ladder} These three guarantees are independent. A permanently bound but incompatible embedding/index pair separates $S1$ from $C1$; advancing between compatible policy/model pairs separates $C1$ from $S1$. A fresh child with its own stable, compatible resources separates $S1+C1$ from $I1$, while propagating live, unchecked resolution shows that $I1$ implies neither $S1$ nor $C1$. The profiles therefore form a partial order: Repeatable Semantic Read provides stability without compatibility, while Compatible Read provides compatibility without stability, so neither dominates the other.
\end{sloppypar}

\begin{sloppypar}\ph{Named isolation levels}
Semantic Read Committed (SRC), $\langle S0,C0,I0\rangle$, favors freshness and permits all four anomalies. Repeatable Semantic Read (RSR), $\langle S1,C0,I1\rangle$, preserves bindings but permits compatibility and merge skew. Compatible Read (CR), $\langle S0,C1,I1\rangle$, enforces inherited contracts but permits semantic rereads and same-identity merge skew. Semantic Snapshot Isolation (SSI), $\langle S1,C1,I1\rangle$, preserves one inherited, stable, compatibility-valid semantic cut and prevents all four. Table~\ref{tab:levels} summarizes the levels and principal costs.\end{sloppypar}

\ph{Choosing an isolation level}
Stability sacrifices freshness, requires historical resources, and may abort when one disappears. Compatibility needs trustworthy contracts; inheritance enlarges checkpoints and may reject branches. A research assistant might choose CR for fresh but compatible resources, whereas a compliance report might choose SSI for one coherent basis.

\ph{A semantic cut, not a timestamp} Unlike a database snapshot, a coherent AI environment may not correspond to a single logical timestamp: providers may not share a clock, and simultaneously available versions can be incompatible. Conversely, an older index remains compatible with the embedding model that built it. SSI therefore preserves one stable, dependency-closed, compatibility-valid \emph{semantic cut}, even when its versions were published at different times.

\section{Runtime Design}
\label{sec:runtime}

\begin{figure}[t]
\centering
\includegraphics[width=\columnwidth]{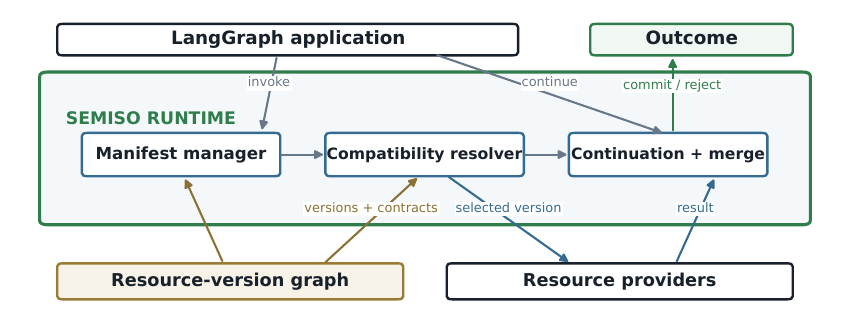}
\Description{The SemIso runtime contains a manifest manager, compatibility resolver, and continuation-and-merge manager. Labeled arrows show application inputs, metadata accesses, provider data accesses, and runtime outcomes.}
\caption{\system's architecture. The manifest manager decides whether to reuse or create a binding; the resolver selects a compatible version or applies the opaque-resource policy; and the continuation manager propagates context and admits or rejects branch unions. Providers remain on the data path and serve only the selected versions.}
\label{fig:runtime}
\end{figure}

\begin{sloppypar}We implement \system as Python middleware for LangGraph 0.6.7; however, the design applies to any durable-workflow framework with checkpoint and continuation hooks. At invocation, the application supplies its input, an isolation level (or custom $S/C/I$ profile), and an abort/capture/taint policy. Independently deployed \emph{providers}---prompt registries, model endpoints, vector indexes, and tool services---register immutable versions and contracts in a shared resource-version graph. \system then uses three modules (Figure~\ref{fig:runtime}): a manifest manager records workflow context, a compatibility resolver selects admissible versions, and a continuation manager propagates context and admits or rejects branch merges.
We discuss each module next.\end{sloppypar}

\ph{Manifest and version graph} The shared graph records each provider version's identity, type, hash, availability, dependencies, and compatibility attributes. The checkpointed per-workflow manifest records its isolation profile, identity-to-version bindings, parent/branch identifiers, decisions, and capture/taint status. Thus the manifest grows as the workflow discovers resources, while the graph changes as owners publish or retire versions.
Owners or applications declare contracts such as ``index $I_1$ requires embedding model $E_1$''; \system enforces but does not infer them.

\ph{Compatibility resolver} For a requested identity, we reuse its manifest binding when stability is required. Otherwise, we scan available versions newest-first and select the first that preserves all dependencies and contracts; we then bind it and add its exact dependencies. A resource discovered after execution begins thereby extends the inherited context. If no version qualifies, the opaque-resource policy applies.

\ph{Continuation manager} The manager stores the profile and manifest in LangGraph's checkpoint.
Retries and resumes restore it; under $I1$, children inherit it and forks receive copies. Before releasing merged output, the manager unions branch manifests and rejects conflicting bindings, violated contracts, unavailable dependencies, or disallowed taint. Rejection preserves the parent checkpoint for recomputation or an explicit weaker policy.

 \ph{Opaque-resource policy} This transaction input governs requests that the resolver cannot bind immutably. \emph{Abort} ends the transaction; \emph{capture} records the observed request/response as a local artifact; and \emph{taint} permits progress while marking the weakened guarantee. Capture preserves the observation, not hidden behavior or side effects.

\system's current prototype supports SRC, RSR, CR, SSI, and caller-specified profiles. In Section~\ref{sec:evaluation}, we use Qdrant 1.16.1 in local mode for real vector search without a separate server.

\section{Evidence and Evaluation}
\label{sec:evaluation}

Our experiments answer three questions:
\begin{enumerate}
\item Do public codebases exhibit conditions that enable semantic anomalies (Section~\ref{sec:empirical-findings})?
\item Can the four anomalies arise independently in a real durable-workflow framework, and which mechanisms prevent them (Section~\ref{sec:system-evaluation})?
\item What metadata cost does \system add (Section~\ref{sec:system-evaluation})?
\end{enumerate}

\subsection{Empirical Findings}
\label{sec:empirical-findings}
\ph{Methodology} We search GitHub repository descriptions and READMEs for ``LangGraph,'' restrict results to Python repositories pushed since January 1, 2025, exclude forks and archived repositories, and sort by stars. The query returns \SearchUniverseN{} repositories. We inspect each repository's latest default-branch HEAD and call it \emph{eligible} if it contains executable Python code that imports LangGraph; \CorpusN{} repositories qualify. Our primary population is the \TopHundredEligible{} eligible codebases with the most stars, of which \TopHundredDurable{} use durable execution. We classify a codebase as \emph{exposed} only when the same executable workflow or directly connected runtime combines durable execution with a mutable or late-bound semantic resource and has no source-visible immutable binding. We manually verify every positive against this same-workflow criterion.

\ph{Findings} Among the top \TopHundredEligible{} codebases, \TopHundredExposed{} meet the exposure rubric: \TopHundredOverallPct\% of the full population and \TopHundredPct\% of its \TopHundredDurable{} durable codebases. These source-visible conditions establish that the risk can arise in real code; they do not show that an anomaly occurred, and controls outside a repository may prevent it.

\begin{figure}[t]
\centering\includegraphics[width=\columnwidth]{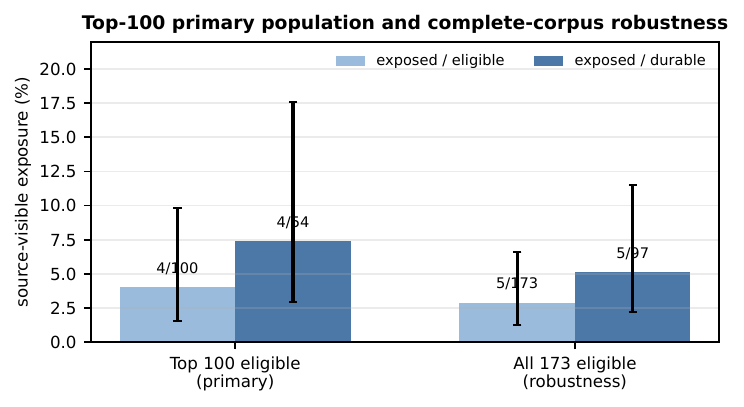}
\Description{Paired bars report exposed over all eligible and exposed over durable codebases with Wilson intervals.}
\caption{Source-visible exposure. The primary top-100 population contains \TopHundredExposed{}/\TopHundredEligible{} exposed codebases overall and \TopHundredExposed{}/\TopHundredDurable{} among durable codebases. The complete search universe is shown as a robustness check; bars include 95\% confidence intervals.}
\label{fig:exposure}
\end{figure}

\subsection{System Evaluation}
\label{sec:system-evaluation}

\ph{Setup} We implement the vendor-risk workflow from Figure~\ref{fig:history} using LangGraph persistence, \texttt{interrupt} and resume, a compiled child subgraph, parallel branches, and Qdrant retrieval. Four focused graphs isolate semantic read skew, compatibility skew, context escape, and merge skew. Deterministic adapters and a fixed deployment schedule provide an oracle for version selection. The two embedding versions have equal dimensions: querying $\mathsf{Emb}_2$ against $\mathsf{Idx}_1$ executes successfully but retrieves the wrong controlled document. Thus the experiments measure framework behavior and enforcement independently of model quality.

\begin{table}[t]
\centering\footnotesize
\setlength{\tabcolsep}{4pt}
\caption{Baseline outcomes: O=occurred/allowed, D/A=detected after commit, P=prevented, N=did not occur.}
\label{tab:outcomes}
\begin{tabular}{lcccc}\toprule
Mechanism & Read & Compat. & Escape & Merge\\\midrule
Default & O & O & O & O \\
Provenance & D/A & D/A & D/A & D/A \\
flow\_version & O & O & O & O \\
Static pinning & P & P & P & P \\
\system (SSI) & P & P & P & P \\
\bottomrule\end{tabular}

\end{table}

\ph{Anomaly isolation} Table~\ref{tab:outcomes} compares five mechanisms. \emph{Default} admits all four anomalies; \emph{Provenance} observes the same histories and reports them only after commit. \texttt{flow\_version} preserves v1 graph-node selection after v2 is deployed, but does not bind external resources. \emph{Static pinning} records every startup-known identity and prevents all four closed-world histories. \system under SSI also prevents all four and rejects the invalid branch union before commit. Separately, runs of the named isolation profiles match Table~\ref{tab:levels}: SRC admits all four anomalies, RSR admits compatibility skew, CR admits semantic rereads and same-identity merge skew, and SSI admits none.

\ph{Open-world discovery} A separate control delays tool selection until approved evidence chooses one of two identities. Static pinning succeeds when both candidates are declared initially; when the candidate universe is open, the selected identity is unbound and resolves to an incompatible live implementation. \system under SSI instead extends the inherited tool-schema constraint when the identity becomes known and selects the compatible retained implementation. Static pinning therefore covers closed dependency sets, while online extension covers discovery during execution.

\ph{Metadata cost} At the largest tested configuration (128 identities and 16 branches), 300 repetitions give p95 resolution latency of \ResolveP{}~$\mu$s and p95 merge-validation latency of \MergeP{}~$\mu$s. A compact-JSON checkpoint with \BoundResources{} bindings and \BranchRecords{} branch records adds \CheckpointDelta{} bytes over the same workflow state without semantic metadata; its root manifest occupies \ManifestBytes{} bytes. Across 30 complete runs per mechanism, the median end-to-end difference remains within timer noise. These laptop measurements cover metadata processing, not the operational cost of retaining historical resource versions.

\section{Related Work}
\label{sec:related}

Database isolation levels~\cite{berenson95,adya99} characterize admitted anomalies. COPS-GT~\cite{cops} gives causal visibility; RAMP~\cite{ramp} gives read-atomic visibility. Brayner et al.~\cite{brayner99} use object semantics to admit nonserial schedules under \emph{semantic serializability}. Earlier transaction terminology differs from ours: Huang et al.~\cite{huang02} use \emph{semantic transactions} for semantics-aware mobile-database scheduling, whereas Cordon~\cite{cordon} uses one to stage and validate agent tool effects. Our \emph{semantic isolation} instead constrains resource versions and compatibility within one durable execution; it neither relaxes schedules nor stages effects. Atomix~\cite{atomix} coordinates tool calls, Mnemosyne~\cite{mnemosyne} repairs generated actions, and ATCC~\cite{atcc} controls concurrent agent-generated SQL. Styx~\cite{styx} provides serializable cloud workflows, while AC/DC~\cite{acdc} extends correctness and recovery across durable workflows.

Temporal~\cite{temporalversion} can pin an execution to a worker deployment version; LangGraph~\cite{langgraphcompat} recommends storing graph-logic choices in application state. These mechanisms protect code paths, not independently managed prompts, model aliases, indexes, or tools. DVC lockfiles~\cite{dvc} and ZenML snapshots~\cite{zenml} bind declared pipelines, while GenericVC~\cite{genericvc} unifies Git-like branching with MVCC and application-specific reconciliation. ML Metadata~\cite{mlmd} records lineage, while MLflow registries~\cite{mlflow} provide addressable versions. MLCask~\cite{mlcask} branches and merges versioned ML pipelines; SPEAR~\cite{spear} makes prompts structured, versioned runtime objects. Qdrant's migration procedure~\cite{qdrantmigration} captures one embedding--index contract; Confining Nondeterminism~\cite{confining} instead places AI-generated artifacts in deterministic versioned dataflow. We combine stable bindings, cross-resource compatibility, continuation inheritance, online extension, and branch admission within one durable execution.

\section{Conclusion and Future Work}
\label{sec:conclusion}

Checkpointing preserves a durable AI workflow's control state, but not necessarily its semantic environment. We formalize this gap through four anomalies and isolation profiles over resource stability, cross-resource compatibility, and continuation inheritance. \system enforces these profiles through online manifest extension, compatible resolution, inherited continuation context, and branch admission. Our repository audit finds the enabling conditions in public LangGraph codebases. We therefore argue that semantic isolation must become an explicit framework contract rather than an application-specific convention.

\ph{Future work: AI workflow serializability}
Semantic Snapshot Isolation is sufficient when workflows only read immutable semantic resources. Workflows that also update shared memory or artifacts, modify semantic configuration, publish outputs consumed concurrently, or perform visible tool actions require a stronger contract. We call a history \emph{AI-workflow serializable} when it is observationally equivalent to a serial ordering of workflows and deployments that preserves program order, resource contracts, shared writes, and visible actions. This prospective property differs from prior \emph{semantic serializability}, which uses database-object semantics to admit otherwise nonserial schedules~\cite{brayner99}. Enforcing AI-workflow serializability requires coordination over writes and effects, not merely version selection.
We leave this extension to future work.

\ph{AI Disclaimer}
The author has used AI tools to polish and shorten portions of this manuscript, but takes
 full responsibility for the ideas and statements therein.

\setlength{\bibsep}{0pt}
\bibliographystyle{ACM-Reference-Format}
\bibliography{references}
\end{document}